\documentclass[galaxies,article,accept,moreauthors,pdftex,10pt,a4paper]{mdpi}
\firstpage{1}
\articlenumber{x}
\doinum{10.3390/------}
\pubvolume{4}
\pubyear{2016}
\copyrightyear{2016}
\externaleditor{Academic Editors: Jose L. G\'{o}mez, Alan P. Marscher and Svetlana G. Jorstad}
\history{Received: 30 August 2016; Accepted: 19 November 2016; Published: date}

\newcommand{\apj}{Astrophys. J.}

\newcommand{\mnras}{Mon. Not. Roy. Astron. Soc.}
\newcommand{\aap}{Astron. Astrophys.}

\newcommand{\enzo}{{\it {\small ENZO}}}
\newcommand{\dd}{\mathrm{d}}
\newcommand{\Mpc}{\mathrm{Mpc}}
\newcommand{\Msun}{\mathrm{M}_{\odot}}

\newcommand{\kpc}{\mathrm{kpc}}
\newcommand{\radio}{\mathrm{radio}}
\newcommand{\ph}{\mathrm{ph}}
\newcommand{\cm}{\mathrm{cm}}

\newcommand{\sek}{\mathrm{s}}
\newcommand{\keV}{\mathrm{keV}}
\newcommand{\MeV}{\mathrm{MeV}}

\newcommand{\grav}{\mathrm{G}}

\newcommand{\erg}{\mathrm{erg}}
\newcommand{\Hz}{\mathrm{Hz}}

\newcommand{\GHz}{\mathrm{GHz}}

\newcommand{\pre}{\mathrm{pre}}
\newcommand{\post}{\mathrm{post}}

\usepackage{microtype}
\usepackage{soul}
 \theoremstyle{mdpi}
 \newcounter{thm}
 \newcounter{ex}
 \newcounter{re}
\Title{Studying the Effect of Shock Obliquity on the $\gamma$-ray and Diffuse Radio Emission in Galaxy Clusters}

\Author{{Denis Wittor}$^{1,*}$, {Franco Vazza}$^{1}$ and {Marcus Br\"{u}ggen}$^1$}

\AuthorNames{Denis Wittor, Franco Vazza, Marcus Br\"{u}ggen}

\address[1]{%
 $^{1}$ \quad Hamburger Sternwarte, Gojenbergsweg 112, 21029 Hamburg, Germany; {dwittor@hs.uni-hamburg.de (D.W.); franco.vazza@hs.uni-hamburg.de (F.V.); mbrueggen@hs.uni-hamburg.de (M.B.)}
 }

\corres{Correspondence: dwittor@hs.uni-hamburg.de; Tel.: +49-040-42838-8485}

\abstract{Observations of diffuse radio emission in galaxy clusters indicate that cosmic-ray electrons are accelerated on $\sim \ \Mpc$ scales. However, protons appear to be accelerated less efficiently since their associated hadronic $\gamma$-ray emission has not yet been detected. Inspired by recent particle-in-cell simulations, we study the cosmic-ray production and its signatures under the hypothesis that the efficiency of shock acceleration depends on the Mach number and on the shock obliquity. For this purpose, we combine \enzo \ cosmological magneto-hydrodynamical simulations with a Lagrangian tracer code to follow the properties of the cosmic rays. Our simulations suggest that the distribution of obliquities in galaxy clusters is random to first order. Quasi-perpendicular shocks are able to accelerate cosmic-ray electrons to the energies needed to produce observable radio emission. However, the $\gamma$-ray emission  is lowered by a factor of a few, $\sim$3 , if cosmic-ray protons are only accelerated by quasi-parallel shocks, reducing (yet not entirely solving) the tension with the non-detection of hadronic $\gamma$-ray emission by the \textit{Fermi}-satellite.}


\begin{document}
%
%
%
\section{Introduction}
 The existence of peripheral, elongated and often polarised radio emission in galaxy clusters, so-called radio relics, gives evidence of cosmic-ray electrons being accelerated by shock waves in the intracluster medium (see \citep{Brunetti_Jones_2014_CR_in_GC} and references therein). Cosmic-ray protons should undergo the same acceleration mechanism, but no evidence of their presence has been found yet. The \textit{Large Area Telescope} on-board of the \textit{Fermi}-satellite \citep{2009ApJ...697.1071A} is searching for signatures of the cosmic-ray protons, which should produce $\gamma$-ray emission through collisions with the thermal gas. Yet no detection of these $\gamma$-rays has been confirmed and for a variety of clusters the upper flux limits have been estimated to be in the range of $0.5-2.2 \times 10^{-10} \ \ph/\sek/\cm^{2}$ above $500 \ \MeV$ \cite{2014ApJ78718A}. Extended searches for the $\gamma$-ray emission in the Coma cluster \cite{2016ApJ819149A} and the Virgo cluster \cite{2015ApJ812159A} have set the flux limits above $100 \ \MeV$ to $5.2 \times 10^{-9}  \ \ph / \sek / \cm^{2}$ for the former and to $1.2 \times 10^{-8}  \ \ph / \sek / \cm^{2}$ for the latter.

 Recent results from particle-in-cell simulations \cite{Caprioli_Spitkovsky_2014_ion_accel_I_eff,Guo_eta_al_2014_I,Guo_eta_al_2014_II} suggest that the efficiency of shock acceleration does not only depend on the shock strength but also on the shock obliquity, e.g., the angle between the shock normal and the underlying upstream magnetic field. Cosmic-ray protons should be accelerated more efficiently by diffusive shock acceleration (DSA) in parallel shocks \cite{Caprioli_Spitkovsky_2014_ion_accel_I_eff}. In contrast, cosmic-ray electrons should prefer a perpendicular configuration as they are first accelerated by shock drift acceleration before they are injected into the DSA cycle \citep{Guo_eta_al_2014_I,Guo_eta_al_2014_II}.

 In our recent work, we have tested if, in galaxy clusters, the additional dependence on the shock obliquity can explain the missing $\gamma$-ray emission and still produce detectable radio relics \citep{2016arXiv161005305W}. In this contribution we present the most relevant results from that work and include new results.
%
%
\section{Methods}
\vspace{-6pt}
\subsection{Cosmological MHD Simulations}
The cosmological magneto-hydrodynamical (MHD) simulation presented in this work has been carried out with the \enzo-code \cite{ENZO_2014}.
In our simulation, we solve the MHD equations \mbox{(see Section 2.1 in \citep{ENZO_2014})} using a piecewise linear method \citep{1985JCoPh..59..264C} in combination with hyperbolic Dedner cleaning \citep{2002JCoPh.175..645D}.
We re-simulate a single galaxy cluster with a final mass of $M_{200}(z\approx 0)\approx 9.74 \times 10^{14} \ \Msun$. The cluster shows a major merger at  $z \approx 0.27$, which is strong enough to produce detectable radio relics (see \citep{2016arXiv161005305W} for further information).
We simulate a $250^3 \ \sim \ \Mpc^3$ comoving volume from $z \approx 30$ to $z \approx 0$, starting from a root grid of $256^3$ cells and $256^3$ dark matter particles. Furthermore, using five levels of Adaptive Mesh Refinement (AMR), we refine $2^5$ times a $\approx$ $25^3$ $\Mpc^3$ region centred around our massive cluster, resulting in a final resolution of $31.7 \ \kpc$ for a large portion of the cluster volume. For the seeding of the large scale-magnetic fields, we use a primordial magnetic seed field with a comoving value of $B_0 = 10^{-10}\ \mathrm{G}$ along each direction.
\subsection{Lagrangian Analysis}
 We track the evolution of cosmic rays using Lagrangian tracer particles (see \citep{2016arXiv161005305W} for more~details). The tracer particles follow, both, the advection of the baryonic matter and the enrichment of shock-injected cosmic rays in time. In post-processing, the tracers are advected in a sub-box consisting of $256^3$ cells of the finest grid of the simulation. The sub-box is centred around the mass centre of our galaxy cluster at $z \approx 0$. The tracers are first injected into the box at $z \approx 1$ following the mass distribution of the gas. During run-time, additional tracers are injected from the boundaries following the mass distribution of the entering gas. In total, we generate $N_p  \approx 1.33\times10^7$ tracers with a final mass resolution of $m_{\mathrm{tracer}} \approx  10^{8} \ \Msun$.

 The tracers are advected linearly in time using the velocities at their location: $\boldsymbol{v}=\boldsymbol{\tilde{v}}+\delta \boldsymbol{v}$. Here, $\boldsymbol{\tilde{v}}$ is the interpolated velocity between the neighbouring cells and $\delta \boldsymbol{v}$ (see Equation (1) in \citep{2016arXiv161005305W}) is a correction term to cure for a possible underestimate due to mixing in complex flows (see \cite{Genel_at_al_2014_following_the_flow} for more~details).

 The local gas values are assigned to every tracer and other properties are computed on the fly. Subsequently, we apply a shock-finding method based on the temperature jump between the positions of a tracer at two consecutive timesteps, similar to the method described in \cite{Ryu_et_al_2003_shock_waves_large_scale_universe}.
 Every time a shock is recorded, the Mach number and the corresponding shock obliquity are computed. The latter is calculated using the velocity jump $\Delta \boldsymbol{v} = \boldsymbol{v}_{\post} - \boldsymbol{v}_{\pre}$ between the pre- and post-shock velocity of the~tracer:

 \begin{equation}
  \theta_i = \arccos \left( \frac{ \Delta \boldsymbol{v} \cdot \boldsymbol{B_i}}{\vert \boldsymbol{v}\vert \vert\boldsymbol{B_i}\vert} \right) \label{eq:theta}.
 \end{equation}

 In the equation above, the index $i$ refers to either the pre- or post-shock values. Across each shock, we compute the kinetic energy flux as $F_{\Psi} = 0.5 \cdot  \rho_{\pre} v_{\mathrm{sh}}^3$, where $\rho_{\pre}$ is the pre-shock density and $v_{\mathrm{sh}}$ is the shock velocity. The thermal energy flux, $F_{\mathrm{th}} = \delta(M)  F_{\Psi}$, and cosmic-ray energy flux, $F_{\mathrm{CR}} = \eta(M) F_{\Psi}$, are computed using the acceleration efficiencies $\delta\left(M\right)$ and $\eta\left( M\right)$ given in \cite{2013ApJ...764...95K}. The efficiency, $\eta(M)$, is taken from \citet{2013ApJ...764...95K} and it includes the effects of magnetic field amplification at the shocks and thermal leakage of suprathermal particles. We include (as in \citep{2014MNRAS.439.2662V,2016arXiv161005305W}) the effect of re-acceleration by computing an effective $\eta_{\mathrm{eff}}(M)$ that is interpolated from the acceleration efficiencies of acceleration and re-acceleration given in \citet{2013ApJ...764...95K}.

 We use the formula given in \citep[][]{2007MNRAS.375...77H} to compute the radio emission from the shocked tracers:

 \begin{equation}
  \begin{split}
   \frac{\mathrm{d} P_{\radio} \left( \nu_{\mathrm{obs}} \right) }{\mathrm{d} \nu} &=  \frac{6.4\cdot10^{34} \ \mathrm{erg}}{\sek \cdot \mathrm{Hz}} \frac{A}{\mathrm{Mpc}^2}  \frac{n_e}{10^{-4} \ \cm^{-3}}
   \frac{\xi_e}{0.05}\left( \frac{T_d}{7 \ \keV} \right)^{\frac{3}{2}}  \\
   &\times \left( \frac{\nu_{\mathrm{obs}}}{1.4 \ \GHz} \right)^{-\frac{s}{2}}
   \frac{\left( \frac{B}{\mu \grav} \right)^{1+\frac{s}{2}}}{\left(\frac{B_{\mathrm{CMB}}}{\mu \grav} \right)^{2}+ \left(\frac{B}{\mu \grav} \right)^{2}} \cdot \eta \left( M \right)
   \end{split}.\label{eq:HB}
 \end{equation}

 The quantities that are taken from the grid are: $A$ the surface area of a tracer, $n_e$ the number density of electrons, $T_d$ the downstream temperature, $B$ the magnetic field strength and the acceleration efficiency $\eta \left(M\right)$ depending on the Mach number $M$ taken from \cite{2013ApJ...764...95K}. We notice that the application of $\eta (M)$ to Equation (\ref{eq:HB}) is limited to spectra flatter than $s \approx -3$, because Equation (\ref{eq:HB}) has been derived in energy space while $\eta(M)$ has been derived in momentum space. However, our modelling is accurate enough for the radio frequency, produced by electrons with Lorentz factor of $\gamma > 10^3$, that we are investigating here \cite{2007MNRAS.375...77H}. The other quantities are the observed frequency band, $\nu_{\mathrm{obs}}=1.4 \ \GHz$, the equivalent magnetic field of the cosmic microwave background, $B_{\mathrm{CMB}} = 3.2 \cdot (1+z)^2 \ \mu \grav$ and electron-to-proton ratio $\xi_e = 0.01$. Following \citet{2007MNRAS.375...77H} we assume that the minimum electron energy is $10$ times the thermal gas energy, while the minimum proton energy is fixed to $780 \ \MeV$.
 We use the same approach as in \cite{2015MNRAS.451.2198V,2010MNRAS.407.1565D,2013A&A...560A..64H} to compute the $\gamma$-ray emission. We refer to Appendix C  of our previous publication \cite{2016arXiv161005305W} for a summary of the method.
%
%
\section{Results}
 In \cite{2016arXiv161005305W} we studied how linking the shock acceleration efficiency to the shock obliquity can affect the acceleration of cosmic rays by predicting the amount of radio and $\gamma$-ray emission produced by either quasi-perpendicular or quasi-parallel shocks. Following {Figure} 3 of \citet{Caprioli_Spitkovsky_2014_ion_accel_I_eff} we define quasi-perpendicular shocks as $\theta \in [50^{\circ}, 130^{\circ}]$ and quasi-parallel shocks as $\theta \in [0^{\circ}, 50^{\circ}] \ \& \ [130^{\circ}, 180^{\circ}]$. However, a more detailed analysis in \citet{2016arXiv161005305W} showed that the effects on the acceleration of cosmic rays are not very sensitive to the selection of $\theta$. \\
 We found that the distribution of shock obliquities in a galaxy cluster roughly follows the distribution of random angles in three-dimensional space, $\propto \sin(\alpha)$. Just based on this, one can expect to observe more quasi-perpendicular shocks than quasi-parallel shocks. Hence, the acceleration of cosmic-ray electrons should be more favoured than the acceleration of cosmic-ray protons. For the results on how this affects the radio and $\gamma$-ray emission, we point to our previous publication \cite{2016arXiv161005305W} as they are similar to the ones presented below.

\begin{figure}[H]
 \centering
  \includegraphics[width = 0.75	\textwidth]{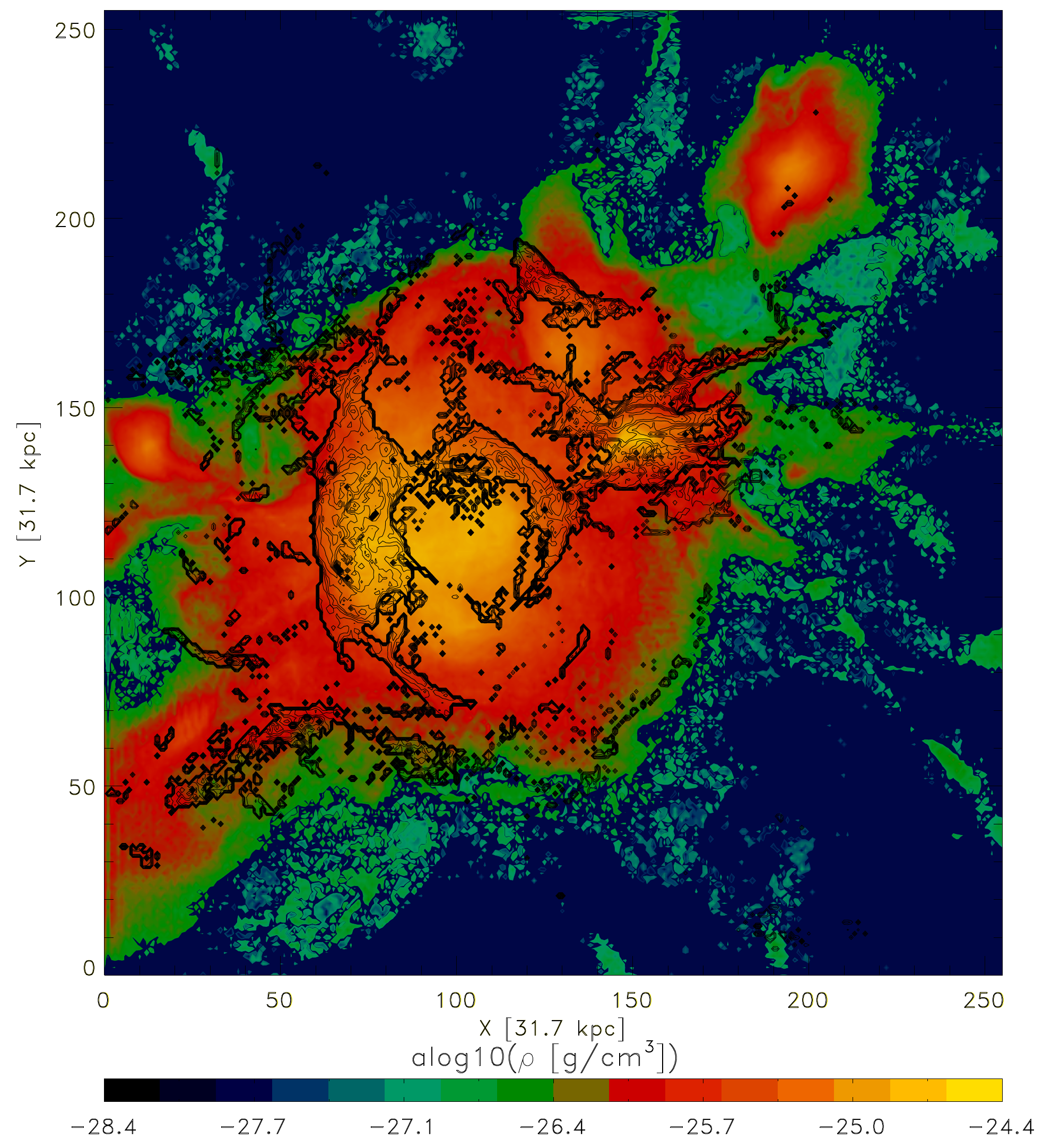}
  \caption{Projected gas density (colours) and radio contours at $z \approx 0.2$. Two radio relics can be seen on the right ($P_{\radio} \approx 3.42 \times 10^{31} \ \erg / \sek / \Hz$) and left ($P_{\radio} \approx 2.26 \times 10^{32} \ \erg / \sek / \Hz$) side of the cluster~centre.}
  \label{subfig:denspRadio}
 \end{figure}

 In this contribution we present a closer analysis of the same cluster at the epoch of the peak of the total radio emission. We show the projection of the gas density overlayed with the radio contours in Figure  \ref{subfig:denspRadio}. The cluster is still in a very active phase after it experienced a major merger at $z \approx 0.27$, and several smaller gas clumps are still falling onto the cluster.

\begin{figure}[H]
\centering
  \includegraphics[width = 0.49\textwidth]{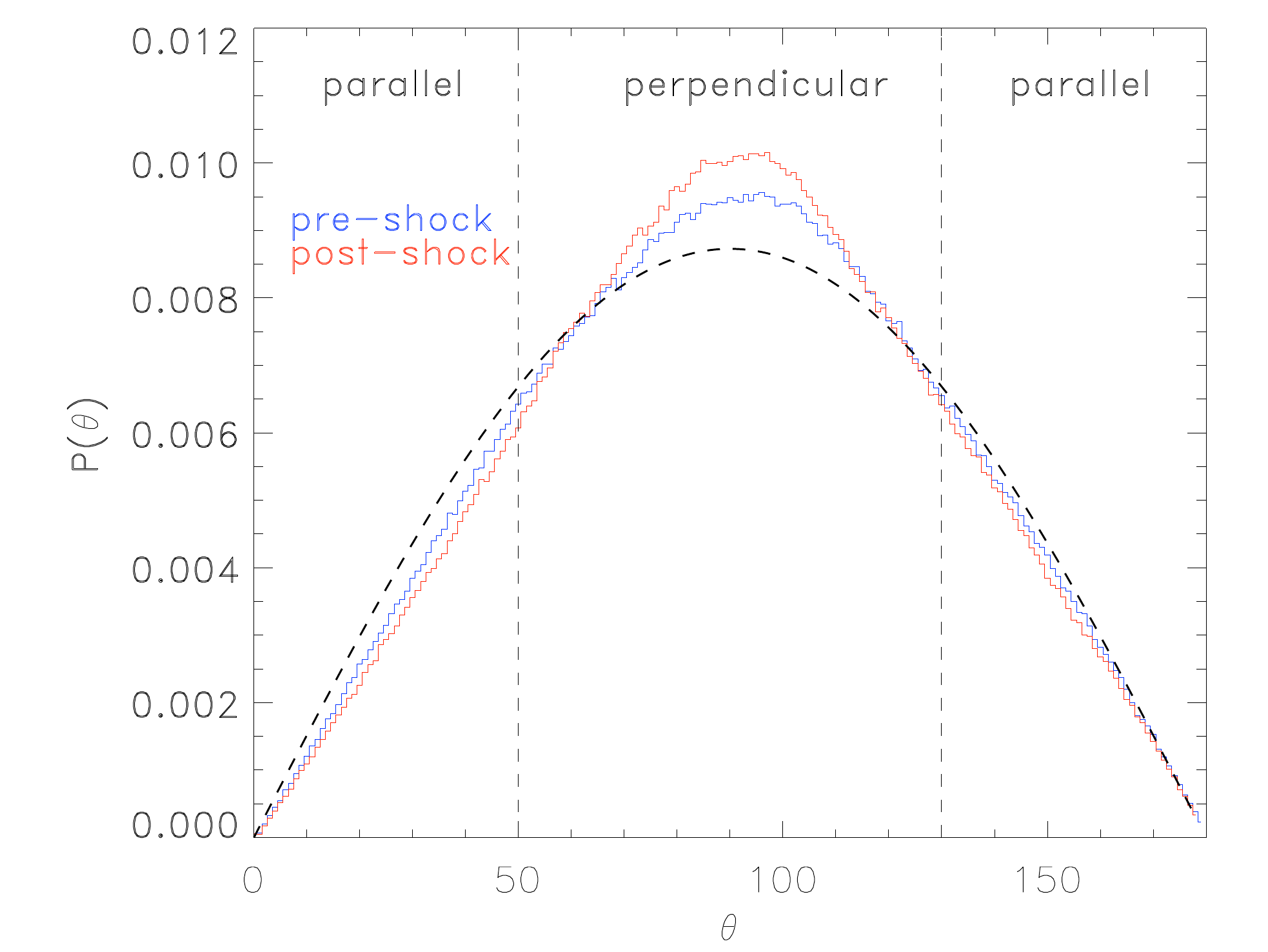}
  \includegraphics[width = 0.49\textwidth]{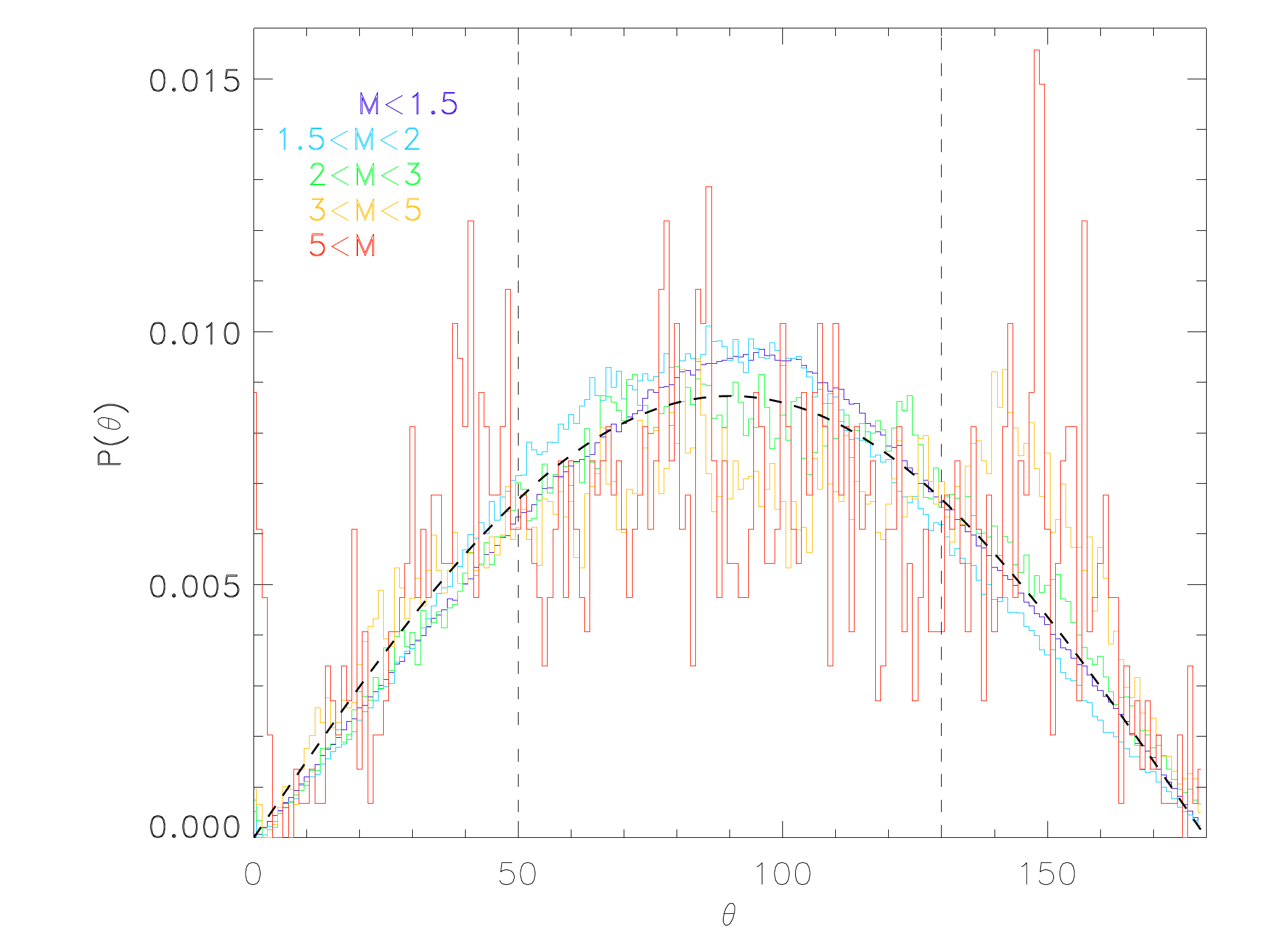}
  \caption{Distributions of shock obliquities at $z \approx 0.2$. The left panel shows the pre-shock (\textbf{blue}) and post-shock (\textbf{red}) distribution of obliquities. The black dashed line shows the expected $\propto sin(\alpha)$ distribution of angles based on pure geometry.  The right panel shows the distribution of pre-shock obliquities for different ranges of Mach numbers: $M < 1.5$ (\textbf{dark blue}), $1.5<M<2$ (\textbf{light blue}), $2<M<3$ (\textbf{green}), $3<M<5$ (\textbf{orange}) and $M>5$ (\textbf{red}).}
  \label{subfig:obliHist}
 \end{figure}

 First, we measure the distribution of pre- and post-shock obliquities at $z \approx 0.2$. The left panel of Figure \ref{subfig:obliHist} shows the measured distributions consistent with isotropy. The right panel of Figure  \ref{subfig:obliHist} shows the distribution of pre-shock obliquities for different selections in the shock Mach numbers. While the obliquity distributions of $M<3$ shocks roughly follow the distribution of random angles, stronger (and rarer) shocks are found to cluster at specific obliquity values, related to single large-scale magnetic structures in the cluster volume.

 In the following we perform a similar analysis as in our previous work \cite{2016arXiv161005305W} to investigate how coupling the shock acceleration efficiencies to the shock obliquity affects the $\gamma$-ray emission \ref{sss:gamma} and the radio emission \ref{sss:radio}.
\subsection{$\gamma$-ray Emission}\label{sss:gamma}
\begin{figure}[H]
 \centering
  \includegraphics[width = 0.6\textwidth]{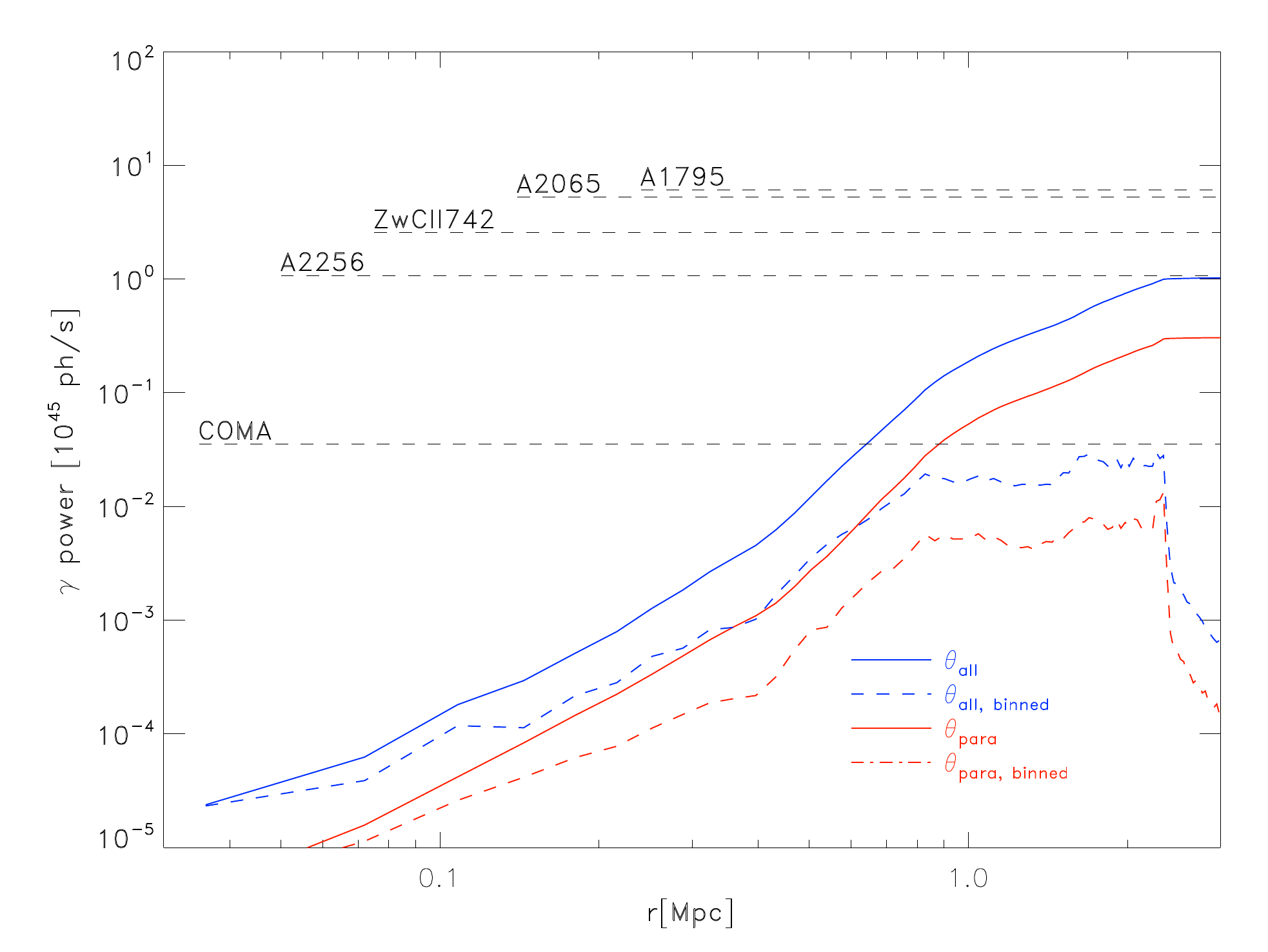} \vspace{-6pt}
  \caption{Profiles of the $\gamma$-ray emission. The solid lines show the total integrated emission profiles and the coloured dashed lines show the radial emission profiles. The $\gamma$-ray emission produced by cosmic-ray protons accelerated in all shocks is given by the blue lines. The red line shows the case of only quasi-parallel shocks being able to accelerate cosmic rays. The black dashed lines show the upper \textit{Fermi}-limits for galaxy clusters that have a comparable mass to our simulated cluster.}
  \label{subfig:gamma}
 \end{figure}

 In Figure \ref{subfig:gamma}, we show the total integrated $\gamma$-ray emission and radial $\gamma$-ray emission profiles produced in the cluster at this epoch. The total $\gamma$-ray emission received from inside $r_{200}$ is $\sim$1.03$\times 10^{45} \ \ph / \sek$, which is above the corresponding \textit{Fermi}-limits (see \cite{2016arXiv161005305W} for the exact computations) of the Coma ($0.035 \times 10^{45}  \ \ph /  \sek$) cluster and just below the limits of A2256 ($1.075 \times 10^{45}  \ \ph  / \sek$). The $\gamma$-ray emission resulting from cosmic-ray protons that have been accelerated only by quasi-parallel shocks is $\sim$0.31$\times 10^{45}~\ \ph / \sek$. This is still above the lowest upper limit of the Coma cluster. The observed drop in $\gamma$-ray emission is consistent with the fact that at low Mach numbers only $\sim$one-third of all shocks are quasi-parallel (see~Figure~\ref{subfig:obliHist}).

 Consistent with our findings from \cite{2016arXiv161005305W}, we conclude that the missing $\gamma$-ray emission cannot be entirely reproduced by limiting the acceleration of cosmic-ray protons to quasi-parallel shocks.
\vspace{-12pt}
\subsection{Radio Emission}\label{sss:radio}
\begin{figure}[H]
 \centering
  \includegraphics[width = 0.32\textwidth]{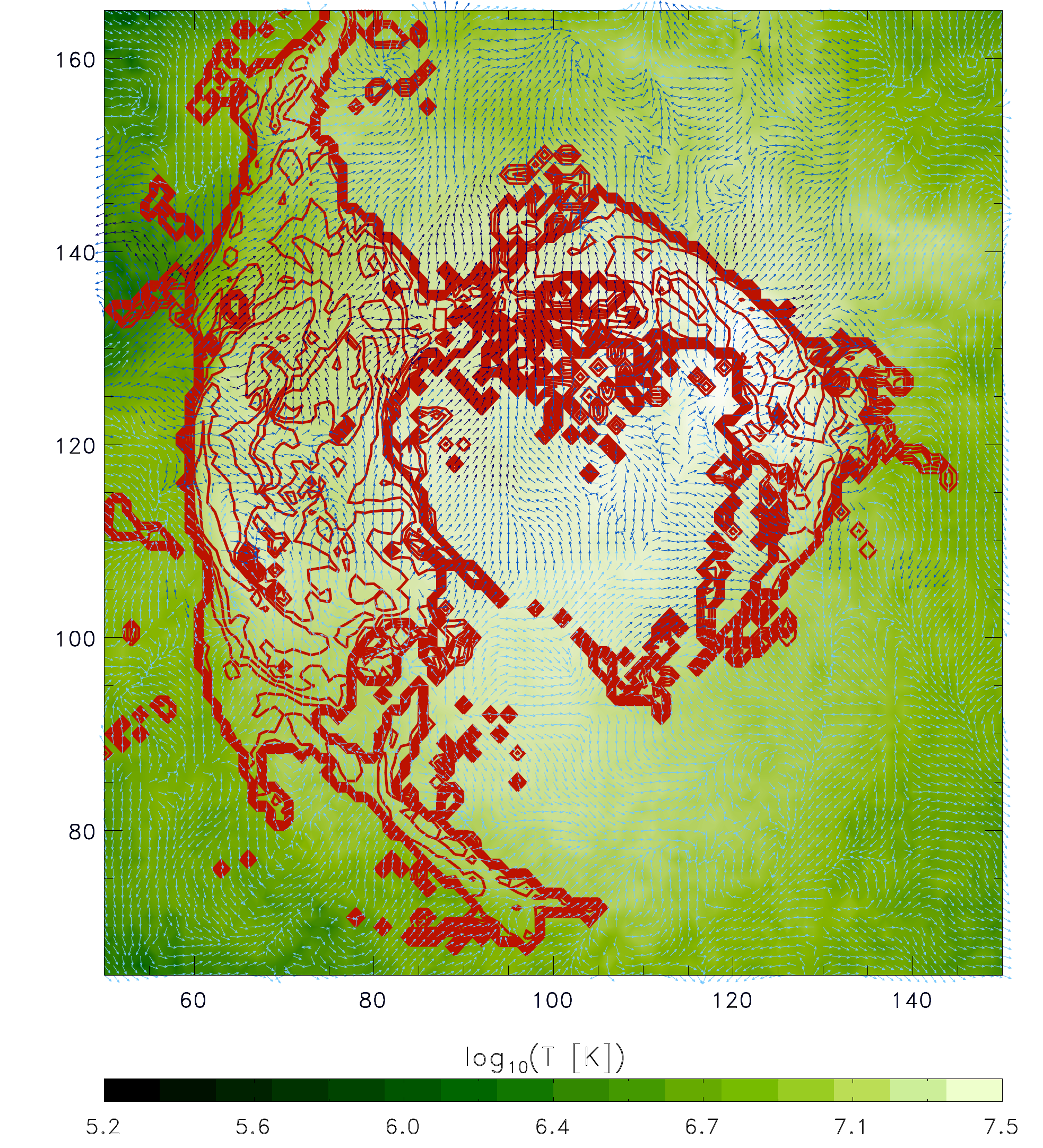}
  \includegraphics[width = 0.32\textwidth]{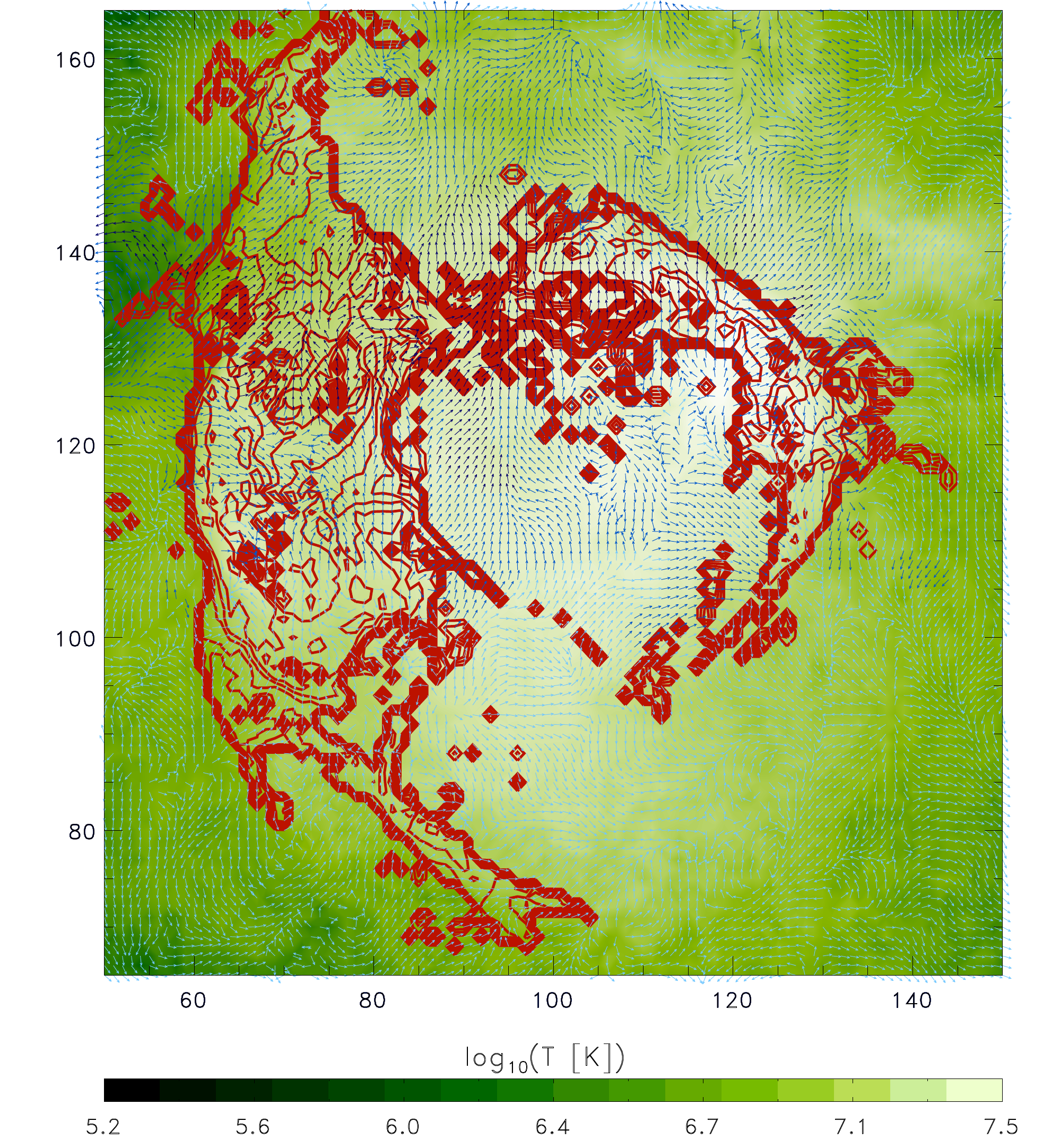}
  \includegraphics[width = 0.32\textwidth]{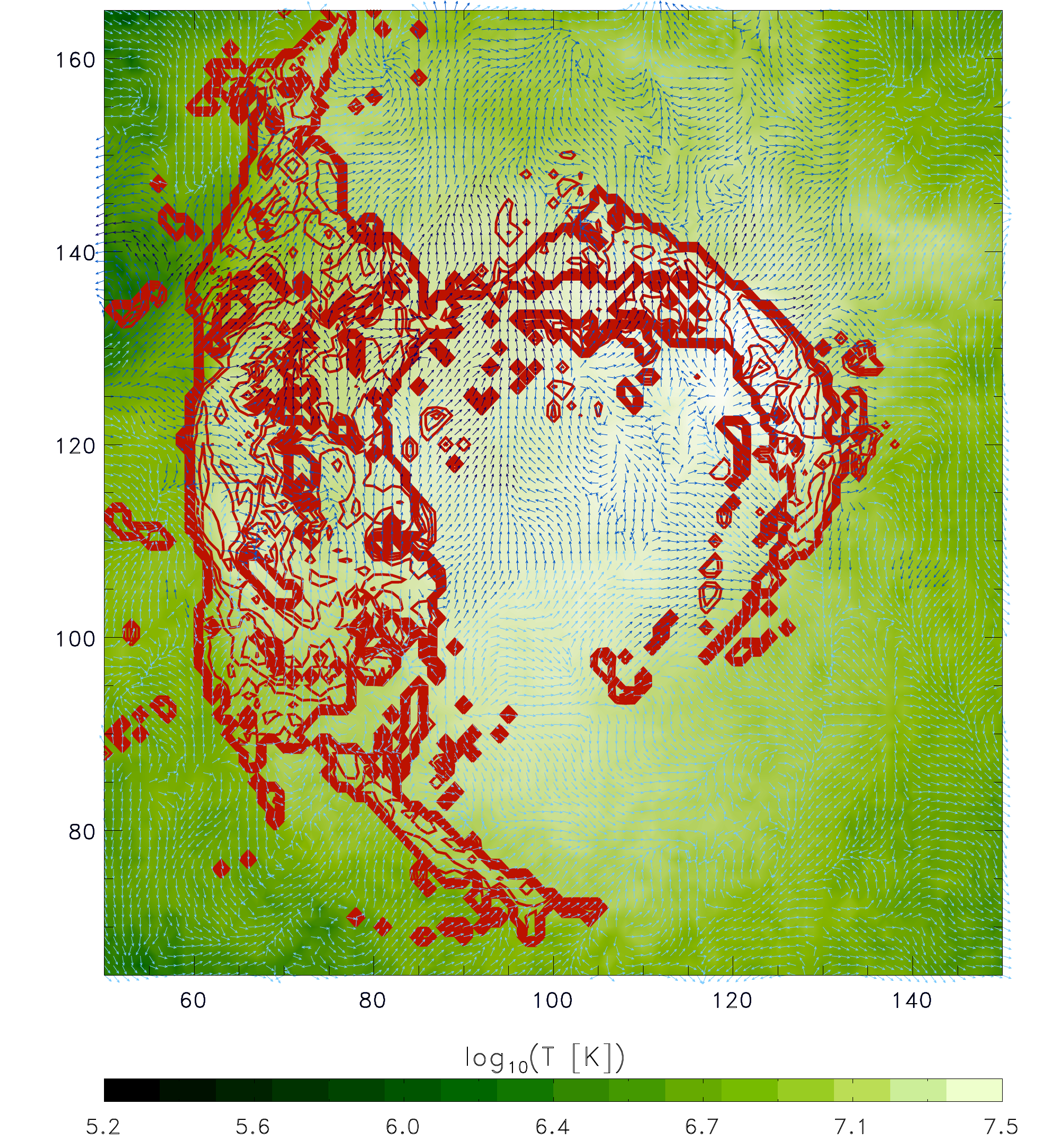}
  \caption{Isolated zoomed versions of our simulated radio relics. The green colours show the gas temperature. The blue arrows show the magnetic field. Their direction indicates the magnetic field direction and the colours give their magnetic field strength using a logarithmical stretching (as the brighter the blue the stronger the magnetic field). The red contours show the radio emission. The left panel shows the relics produced by all cosmic-ray electrons. The middle panel shows the relics produced by electrons that have been accelerated by quasi-perpendicular shocks only. The right panel shows the relics produced by electrons that have been accelerated by quasi-parallel shocks only. The axes are in $\dd x = \ 31.7 \ \kpc$ units.}
  \label{subfig:TnvCCpc}
 \end{figure}

 We observe two radio relics on the left (hereafter relic one) and right (hereafter relic two) side of the cluster core (see Figure \ref{subfig:denspRadio}). Both relics are in the range detectability by modern radio observations. Figure \ref{subfig:TnvCCpc} shows the complex geometry of the magnetic field lines in the relic regions. We observe that the morphologies of the relics do not change significantly, if only either quasi-perpendicular (middle panel) or quasi-parallel (right panel) shocks are able to accelerate cosmic-ray electrons.

\begin{figure}[H]
 \centering
  \includegraphics[width = 0.49\textwidth]{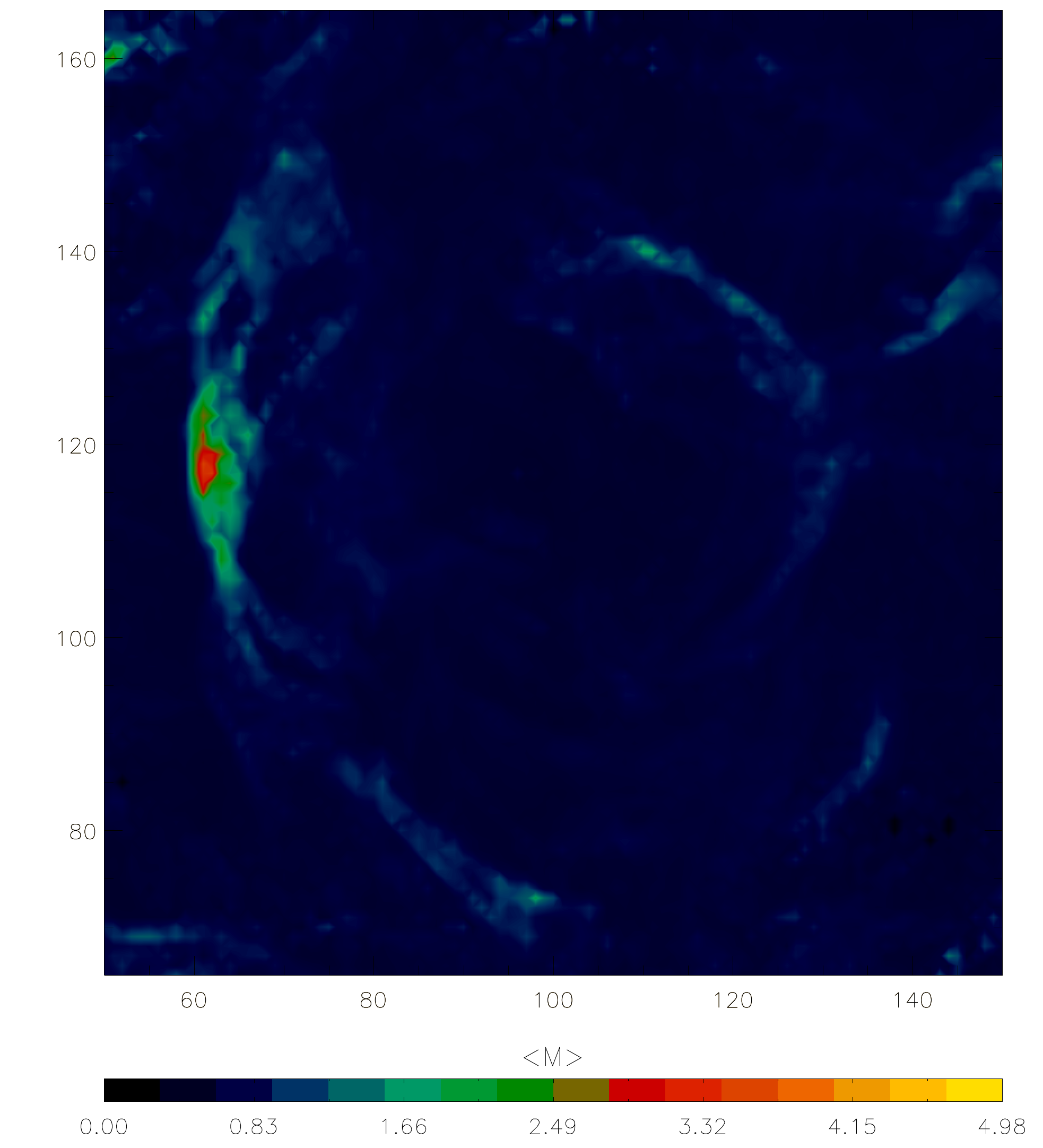}
  \includegraphics[width = 0.49\textwidth]{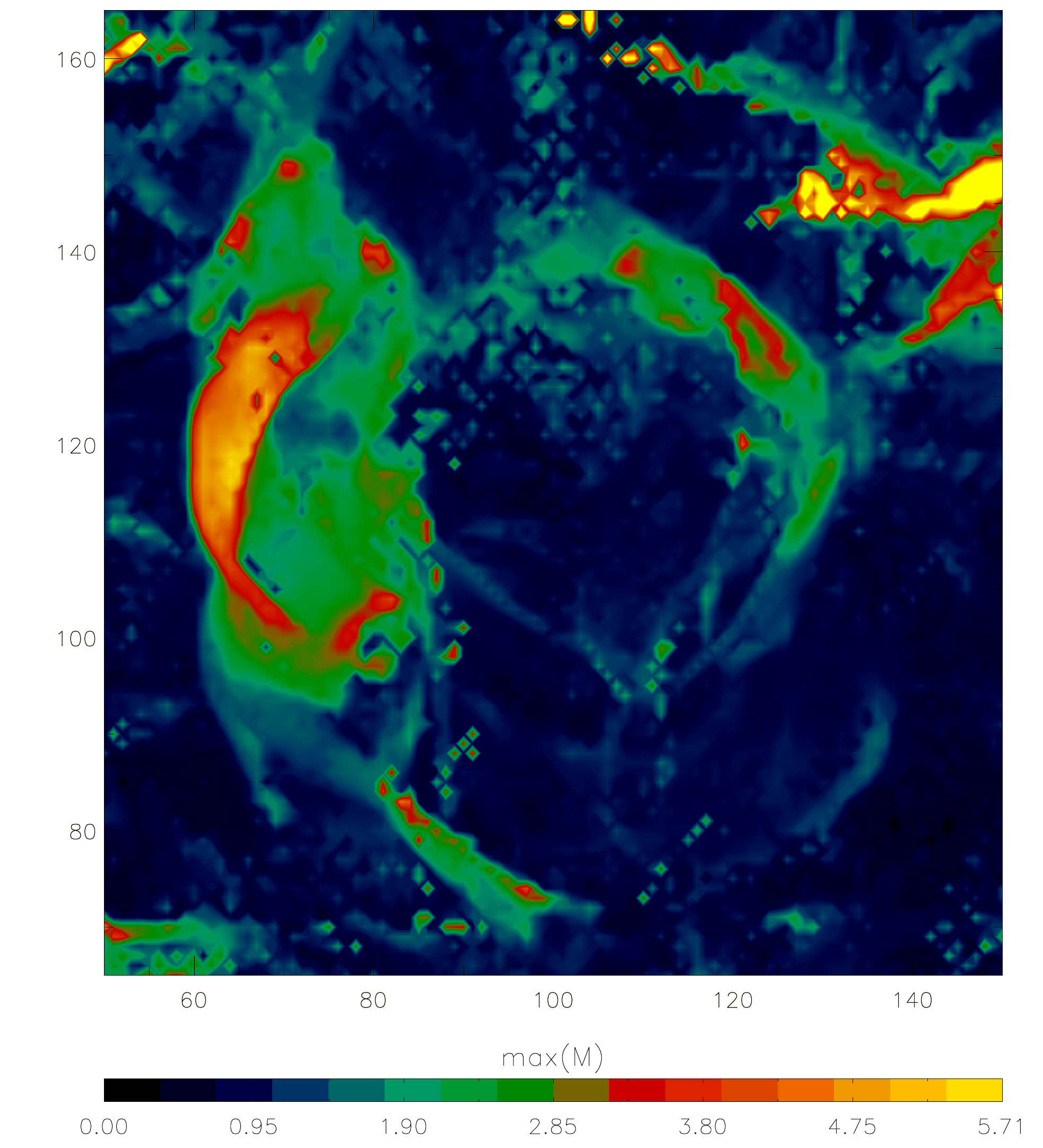}
  \caption{Maps of the mean (\textbf{left} panel) and maximum (\textbf{right} panel) Mach numbers of the shocks that are producing the radio relics. The axes are in $\dd x = \ 31.7 \ \kpc$ units.}
  \label{subfig:mnomask}
 \end{figure}

 In the first relic, only $\sim \ 46 \%$ of the radio emission is produced by cosmic-ray electrons that have been accelerated by quasi-perpendicular shocks. This is consistent with the distribution of obliquities (see Figure \ref{subfig:obliHist}), as relic one is mostly produced by a higher Mach number shocks with $M \ \sim \ 3-5$ (see~Figure~\ref{subfig:mnomask}), that do not follow the distribution of random angles in a three-dimensional space (see~Figure \ref{subfig:obliHist}). On the other hand, $\sim$59$\%$ of the shocks producing relic two are quasi-perpendicular, as it is produced by weaker shocks with $M \ \sim  \ 2-3$ (see~Figure \ref{subfig:mnomask}).

 However, consistent with \cite{2016arXiv161005305W}, we find that both simulated relics remain visible if the acceleration of electrons is limited to quasi-perpendicular shocks.
%
%
%
\section{Discussion}
 Combining MHD-simulations and Lagrangian tracers, we continued our study on how restricting the shock acceleration efficiencies to the obliquity affects cosmic rays in galaxy clusters. At the epoch of the highest radio emission, we examined how cosmic rays, that have been accelerated by either quasi-parallel or quasi-perpendicular shocks, contribute to the resulting $\gamma$-ray and radio emission. We chose this epoch for our investigation as the two radio relics are the most prominent.

 Our findings agree with our results from \cite{2016arXiv161005305W}: The distribution of shock obliquities follows the distribution of random angles in a three-dimensional space. Furthermore, we discovered that this only holds for low Mach numbers $M \le 3$. The distribution of shock obliquities for the few high Mach numbers $M \ge 3$ does not show this trend, as they tend to cluster around single magnetic field structures.

 Consistent with our findings from \cite{2016arXiv161005305W}, the $\gamma$-ray emission drops by a factor of $\sim$3 if only quasi-parallel shocks are able to accelerate the cosmic rays. Yet, this drop is not large enough the explain the low upper limits set by the \textit{Fermi}-satellite \cite{2014ApJ78718A}, especially in the case of the Coma cluster \cite{2016ApJ819149A}.

 On the other hand, the radio emission remains observable if only quasi-perpendicular shocks are able to accelerate cosmic rays. This also holds if the majority of the radio emission is produced by a strong quasi-parallel shock. This supports our conclusion from \cite{2016arXiv161005305W} that it is possible that the cosmic-ray electrons in observed radio relics have only been accelerated by quasi-perpendicular shocks.

 We mention that we do not include any other additional mechanisms, such as cosmic-ray re-acceleration by cluster weather or turbulence (e.g., \citep{2011MNRAS.412..817B}) which would produce further cosmic-ray protons. On the other hand, we do not allow any spatial diffusion of the cosmic-rays, that would reduce the $\gamma$-ray flux through proton accumulation in the cluster outskirts (e.g., \citep{2013MNRAS.434.2209W,2016arXiv160702042L}).

\vspace{6pt}

%
%
%
\acknowledgments{The cosmological simulations accomplished in this work were performed using the \enzo \ code (http://enzo-project.org), and were partially produced at Piz Daint (ETHZ-CSCS, http://www.cscs.ch) in the Chronos project ID ch2 and s585, and on the JURECA supercomputer at  the NIC of the Forschungszentrum J\"{u}lich,  under allocations no. 7006 and 9016 (FV) and 9059 (MB). DW acknowledges support by the Deutsche Forschungsgemeinschaft (DFG) through grants SFB 676 and BR 2026/17. FV acknowledges personal support from the grant VA 876/3-1 from the DFG. FV and MB also acknowledge partial support from the grant FOR1254 from DFG.
We computed all cosmological distances using the Ned Cosmology Calculator \citep{2006PASP..118.1711W}.
We acknowledge fruitful discussions with Tom Jones, Klaus Dolag and Claudio Gheller.}
%
\authorcontributions{D. Wittor produced the analysis in this work and wrote the paper.  All authors performed the simulations used in this work and contributed to their theoretical interpretation.}
%
\conflictofinterests{The authors declare no conflict of interest. The founding sponsors had no role in the design of the study; in the collection, analysis, or interpretation of data; in the writing of the manuscript, and in the decision to publish the results.}
%
%
\bibliographystyle{mdpi}
\renewcommand\bibname{References}

%
%
\end{document}